\documentclass[aps,prd,twocolumn,nofootinbib,showpacs,groupedaddress,preprintnumbers,amsmath,amssymb,floatfix]{revtex4-2}

\usepackage[bookmarks, linktocpage, colorlinks = true, linkcolor = blue, urlcolor  = blue, citecolor = blue, anchorcolor = green, hyperindex = true, hyperfigures]{hyperref}

\usepackage{graphicx} 
\usepackage{amssymb}
\usepackage{dcolumn}
\usepackage{multirow}
\usepackage{amsmath,amssymb}
\usepackage{slashed}
\usepackage[usenames]{color}
\usepackage{float}

\newcommand{\MeV}{\text{MeV}}
\newcommand{\GeV}{\text{GeV}}
\newcommand{\fm}{\text{fm}}
\newcommand{\muB}{\mu_{B}}
\newcommand{\muI}{\mu_{I}}
\newcommand{\nB}{n_{B}}
\newcommand{\nsat}{n_{\mathrm{sat}}}
\newcommand{\gbar}{\bar{g}}
\newcommand{\Nc}{N_{c}}
\newcommand{\Nf}{N_{f}}
\newcommand{\calD}{\mathcal{D}}

\newcommand{\calM}{\mathcal{M}}
\newcommand{\calO}{\mathcal{O}}

\newcommand{\bk}{\boldsymbol{k}}
\newcommand{\bq}{\boldsymbol{q}}
\newcommand{\bkhat}{\hat{\boldsymbol{k}}}
\newcommand{\bgamma}{\boldsymbol{\gamma}}

\newcommand{\Pid}{P_{\mathrm{id}}}
\newcommand{\Lambdabar}{\bar{\Lambda}}
\newcommand{\btilde}{\tilde{b}}
\newcommand{\etilde}{\tilde{\epsilon}}

\DeclareMathOperator{\tr}{tr}
\DeclareMathOperator{\Ai}{Ai}
\DeclareMathOperator{\Bi}{Bi}
\DeclareMathOperator{\SumInt}{
\mathchoice
  {\ooalign{$\textstyle\sum$\cr\hidewidth$\displaystyle\int$\hidewidth\cr}}
  {\ooalign{\raisebox{.14\height}{\scalebox{.7}{$\scriptstyle\sum$}}\cr\hidewidth$\textstyle\int$\hidewidth\cr}}
  {\ooalign{\raisebox{.2\height}{\scalebox{.6}{$\scriptstyle\sum$}}\cr$\scriptstyle\int$\cr}}
  {\ooalign{\raisebox{.2\height}{\scalebox{.6}{$\scriptstyle\sum$}}\cr$\scriptstyle\int$\cr}}
}

\newcommand{\changed}[1]{\underline{#1}}

\begin{document}

\preprint{INT-PUB-23-052}
\title{Enhanced contribution of the pairing gap to the QCD equation of state at large isospin chemical potential}
\author{Yuki~Fujimoto}
\email{yfuji@uw.edu}
\affiliation{Institute for Nuclear Theory, University of Washington, Box 351550, Seattle, WA 98195, USA}

\date{\today}

\begin{abstract}
I study QCD at large isospin density, which is known to be in the superfluid state with Cooper pairs carrying the same quantum number as pions.
I solve the gap equation derived from the perturbation theory up to the next-to-leading order corrections.
The pairing gap at large isospin chemical potential is found to be enhanced compared to the color-superconducting gap at large baryon chemical potential due to the $\sqrt{2}$ difference in the exponent arising from the stronger attraction in one-gluon exchange in the singlet channel.
Then, using the gap function, I evaluate the contribution of the condensation energy of the superfluid state to the QCD equation of state.
At isospin chemical potential of a few GeV, where the lattice QCD and the perturbative QCD can be both applied, the effect of the condensation energy becomes dominant even compared to the next-to-leading order corrections to the pressure in the perturbation theory.
It resolves the discrepancy between the recent lattice QCD results and the perturbative QCD result.
\end{abstract}

\maketitle

\section{Introduction}

The better understanding of QCD at nonzero chemical potential is essential for unraveling the dynamical phenomena involving the strong interaction such as binary neutron star mergers, core-collapse supernovae, and heavy-ion collisions, from the first principles.
The QCD equation of state (EoS) is the most important quantity characterizing the thermodynamics of the system.

The QCD EoS can be evaluated reliably in perturbation theory at small temperature $T$ and large quark chemical potential $\mu$~\cite{Freedman:1976xs, *Freedman:1976dm, *Freedman:1976ub, Baluni:1977ms, Vuorinen:2003fs, Kurkela:2009gj, Gorda:2018gpy, Gorda:2021znl, *Gorda:2021kme,  Gorda:2023mkk}.
The validity range of perturbative QCD (pQCD) is limited to the range $\mu \gtrsim 1~\GeV$.
This value corresponds to the baryon density $\nB \gtrsim 50~\nsat$, where $\nsat \simeq 0.16~\fm^{-3}$ is the nuclear saturation density.
It is far beyond the reach of the core density of heavy neutron stars, so the pQCD cannot be directly applied to the realistic environment although it has an indirect impact~\cite{Komoltsev:2021jzg, Gorda:2022jvk} (see, however, Refs.~\cite{Somasundaram:2022ztm, Zhou:2023zrm}).
In the non-perturbative regime, lattice QCD simulations have pinned down the EoS at high $T$ and small $\mu$ by Taylor expansion in terms of $\mu / T$~\cite{HotQCD:2018pds, Borsanyi:2020fev}.
However, the large-$\mu$ region of QCD at $\mu / T > 1$ is generally inaccessible by the current Monte-Carlo algorithm due to the sign problem (see Ref.~\cite{Nagata:2021ugx} for review).

The sign problem can be circumvented in two-flavor QCD at nonzero chemical potential by taking the same values of chemical potentials with opposite signs for $u$ and $d$ quarks~\cite{Alford:1998sd}.
This specific setup corresponds to setting nonzero chemical potential for the (third component of) \emph{isospin}, which I denote as $\muI$, while keeping the baryon chemical potential, $\muB$, to zero.
There have been several lattice QCD studies of multi-pion system at nonzero $\muI$ regarding the phase structure and thermodynamic properties~\cite{Kogut:2002tm, Kogut:2002zg, Kogut:2004zg, deForcrand:2007uz, Beane:2007qr, Beane:2007es, Detmold:2008gh, Detmold:2008fn, Detmold:2008yn, Detmold:2008bw, Cea:2012ev, Detmold:2012wc, Detmold:2012pi, Endrodi:2014lja, Brandt:2017oyy, Brandt:2018omg, Brandt:2022hwy, Brandt:2023kev, Abbott:2023coj} (see Ref.~\cite{Beane:2010em} for review of some of the earlier works).

A recent lattice QCD calculation~\cite{Abbott:2023coj} provided the QCD EoS at $T\approx 0$ and up to $\muI \simeq 3~\GeV$.
The authors were able to construct states with the quantum numbers of up to 6144 pions, which correspond to large $\muI$, on ensembles of gauge configurations based on an ingenious algorithm introduced in Ref.~\cite{Detmold:2010au, Detmold:2014iha}, and measured the thermodynamic properties from the extracted ground state energies of these systems.
They compared their results with the pQCD calculation from Ref.~\cite{Graf:2015tda}, and they found a discrepancy between their lattice results and the pQCD calculation at $\muI \gtrsim 2~\GeV$.

QCD at nonzero $\muI$ can be regarded as the phase-quenched theory, in which the complex phase of the fermion determinant is neglected, of QCD at nonzero $\muB$.
Recently, Moore and Gorda pointed out that  phase quenching works for any linear combination of chemical potentials not only for $\muI$.
They claim that the relative difference between the phase-quenched theory and the original theory is $\calO(\alpha_s^3)$, where $\alpha_s$ is the QCD coupling constant, from the perturbative consideration~\cite{Moore:2023glb}.
Nevertheless, the perturbative $\calO(\alpha_s^3)$ difference between the phase-quenched theory and the original theory will not account for the discrepancy between the lattice QCD and pQCD calculations mentioned above as the latter discrepancy is noticeably larger than $\calO(\alpha_s^3)$.

Further, Moore and Gorda proposed that the phase-quenched lattice calculation can be used to extract the pressure of original theory up to $\calO(\alpha_s^4)$ corrections in the perturbation theory through the rigorous QCD inequality~\cite{Cohen:2003ut}.
By the use of this inequality, the phase-quenched theory places the bound on the pressure of the original theory (see Ref.~\cite{Fujimoto:2023unl} for an application).

The purpose of this work is to demonstrate that the EoS of the phase-quenched theory has an exponentially enhanced nonperturbative correction compared to the original theory.
To this end, I take QCD at nonzero $\muI$ as an example.
In this specific theory, the nonperturbative correction to the EoS related to the phase quenching is embodied as the BCS condensation energy.

The ground state of the isospin QCD is expected to be a Bose-Einstein condensate (BEC) phase with pion condensation at $\muI > m_\pi$.
As $\muI$ increases, it undergoes crossover to the Bardeen-Cooper-Schrieffer (BCS) state with Cooper pairs carrying the same quantum number as pions~\cite{Son:2000xc, *Son:2000by} as observed in the recent lattice QCD result~\cite{Abbott:2023coj} (see Ref.~\cite{Mannarelli:2019hgn} for a review).
In the isospin QCD, $u$ and $\bar{d}$ quarks fill up the Fermi sea, and form the color singlet quark-antiquark condensate, while in two-flavor QCD at nonzero $\muB$ and zero $\muI$, $u$ and $d$ quarks fill up the Fermi sea, and form the color non-singlet diquark condensate leading to the color superconductivity.
The former ($q\bar{q}$ channel) has stronger one-gluon attraction compared to the latter ($qq$ channel), thus the pairing gap is exponentially enhanced at nonzero $\muI$.

The BCS condensation energy correction to the pQCD calculation also accounts for the discrepancy between the lattice QCD and pQCD results.
The gap parameter in the isospin QCD is evaluated by using a method similar to those used to derive the diquark gap parameter in the color-superconducting phase of QCD at nonzero $\muB$ (see Refs.~\cite{Rischke:2003mt, Alford:2007xm} and reference therein).
I note that the exponential enhancement of the BCS gap~\cite{Son:2000xc, *Son:2000by} and the significance of the condensation energy contribution to the EoS in the pQCD calculation~\cite{Chiba:2023ftg} (see also Ref.~\cite{Kojo:2014rca} for the demonstration that the attractive interaction near the Fermi surface stiffens the EoS) were previously pointed out in the literature although there was no reliable evaluation of these effects prior to this work.

The paper is organized as follows.
In the next section, I review the QCD EoS at nonzero $\mu$.
I review the path integral representation and QCD partition function, including the notion of the phase quenching and the perturbative expansion of the QCD EoS.
In Sec.~\ref{sec:cooper}, I discuss the Cooper pairing in QCD at nonzero $\muI$.
After an overview of the Cooper pairing is given, the derivation and solution of the gap equation are presented.
Section~\ref{sec:cond} constitutes the main result of this paper.
Based on the pairing gap obtained in the previous section, I calculate the condensation energy and show the numerical results.
In Sec.~\ref{sec:qhc}, I mention the possible implications of this work to the quark-hadron crossover.
Finally, I end the paper with a summary and discussion.

Throughout the paper, I mainly follow the notations in Ref.~\cite{Alford:2007xm}, and take the number of flavors $\Nf$ as $\Nf=2$ unless otherwise stated.

\section{Review of QCD equation of state at nonzero chemical potential}

In this section, I first review  the Euclidean path integral representations of the QCD partition function with nonzero chemical potential along with the notion of phase quenching.
Then, I discuss the perturbative expansion of the partition function and the phase-quenching effect in the perturbation theory.
The problem setting in this work and the possible resolution are briefly mentioned.

\subsection{Path integral representation of QCD partition function}

Here, I introduce the notion of phase quenching, as discussed in detail in Refs.~\cite{Cohen:2003ut, Moore:2023glb}.
The Dirac operator $\calD_f(\mu_f)$ for a quark of flavor $f$ at nonzero chemical potential $\mu_f$ is defined as
\begin{equation}
    \calD_f(\mu_f) \equiv \slashed{D} + m_f + \mu_f \gamma^0\,,
\end{equation}
where the covariant derivative, $\slashed{D} \equiv \gamma^\mu \partial_\mu + i g \gamma^\mu T_a A^a_{\mu}$, is a skew-Hermitian operator, i.e. $\slashed{D}^\dagger = -\slashed{D}$.
Integrating out the fermionic degrees of freedom, the grand canonical partition function in the path integral representation is expressed as
\begin{equation}
    Z(T, \{\mu_f\}) = \int [dA] \left( \prod_{f}  \det \calD_f(\mu_f) \right) e^{-S_{\rm G}}\,,
\end{equation}
where $S_{\rm G}$ is the Euclidean action of QCD in the gauge sector.
In general, the fermion determinant $\det \calD_f(\mu_f)$ is complex-valued at nonzero $\mu_f$.
This complex phase is the source of the sign problem preventing the Monte-Carlo simulation on the lattice.

I define the phase-quenched theory by discarding the phase of the fermion determinant as
\begin{equation}
    Z_{\rm PQ} (T, \{\mu_f\}) = \int [dA] \left( \prod_{f}  \big| \det \calD_f(\mu_f) \big| \right) e^{-S_{\rm G}}\,.
\end{equation}
Because the complex phase is absent, the phase-quenched theory is free from the sign problem.
From the relation
\begin{equation}
    \label{eq:phlost}
    \gamma^5 \calD(\mu_f) \gamma^5 = \calD^\dagger(-\mu_f)\,,
\end{equation}
one can rewrite the phase-quenched fermion determinant as
\begin{equation}
    \label{eq:pqdet}
    \big| \det \calD_f(\mu_f) \big| = \sqrt{\det \calD_f(\mu_f) \det \calD_f(-\mu_f)}\,.
\end{equation}
Therefore, the phase-quenched theory is equivalent to a theory with an equal number of fermions with opposite chemical potentials.
The fermion determinant appears with a fractional power $1/2$, so this theory is unphysical.

The phase-quenched theory becomes physically sensible in two flavors with mass-degenerate $u$ and $d$ quarks.
In this theory, $u$ and $d$ quarks have opposite chemical potentials $\mu_u = \mu$ and $\mu_d = - \mu$.
Usually, the quark chemical potential $\mu$ is relabeled as $\muI \equiv 2 \mu$, where $\muI$ is a conjugate variable to the third component of isospin $I_3$ and it is called isospin chemical potential.
The corresponding partition function is defined as
\begin{align}
        Z_{I} (T, \muI) &= \int[dA] \det \calD\left(\frac{\muI}{2} \right) \det \calD\left(-\frac{\muI}{2} \right)  e^{-S_{\rm G}} \notag\\
        &= \int[dA] \left|\det \calD\left(\frac{\muI}{2} \right) \right|^2  e^{-S_{\rm G}}\,.
\end{align}
From the last line, one can indeed verify this is the phase-quenched version of a theory at nonzero baryon chemical potential $\muB \equiv \Nc \mu$ and zero $\muI$:
\begin{equation}
    Z_{B} (T, \muB) = \int[dA] \left[\det \calD\left(\frac{\muB}{\Nc} \right)\right]^2 e^{-S_{\rm G}}\,.
\end{equation}

\subsection{Perturbative expansion}
\label{sec:pert}
Hereafter, instead of the partition function, I consider the pressure $P = (T/V) \ln Z$.
I quote the perturbation expansion of pressure up to next-to-next-to-leading order (NNLO) in the massless limit~\cite{Freedman:1976xs, *Freedman:1976dm, *Freedman:1976ub, Baluni:1977ms} regularized in the $\overline{\rm MS}$ scheme~\cite{Fraga:2001id, Kurkela:2009gj}
\begin{equation}
    \label{eq:ppqcd}
    P = P_{\rm LO} + \alpha_s P_{\rm NLO} + \alpha_s^2 P_{\rm NNLO}\,.
\end{equation}
and the coefficients at each order in $\alpha_s$ are
\begin{align}
    P_{\rm LO} &= \Pid\,, \label{eq:PLO}\\
    P_{\rm NLO} &= - \frac{2}{\pi}\Pid\,, \label{eq:PNLO}\\
    P_{\rm NNLO} &= - \frac{1}{\pi^2} \bigg[\Nf \ln \left(\Nf\frac{\alpha_s}{\pi}\right)
    + \frac{\beta_0}{2} \ln \frac{\Lambdabar^2}{(2\mu)^2} \notag\\
    &\qquad\qquad + 18 - 0.99793 \Nf \bigg] \Pid\,, \label{eq:PNNLO}
\end{align}
where $\beta_0 \equiv (11/3) \Nc - (2/3) \Nf$ is the first coefficient of the QCD beta function with $\Nc$ being the number of colors.
I also define the pressure of the ideal quark gas as
\begin{equation}
    \label{eq:pid}
    \Pid \equiv \Nc \Nf \frac{\mu^4}{12 \pi^2}\,.
\end{equation}
For the coupling constant $\alpha_s(\Lambdabar)$, I use the expression at the NNLO, and take into account the running of $\alpha_s(\Lambdabar)$, which is evaluated at the renormalization scale $\Lambdabar$.
The $\overline{\rm MS}$ scale is fixed as $\Lambda_{\overline{\rm MS}} \simeq 330~\text{MeV}$, which is the value suggested from the $\Nf=2$ lattice-QCD data~\cite{Fritzsch:2012wq, FlavourLatticeAveragingGroupFLAG:2021npn}.
I set $\bar{\Lambda} = 2\mu$ in the following calculation as $2\mu$ is a typical hard interaction scale in the system, but there is an ambiguity in the choice of $\bar{\Lambda}$.
As in the conventional prescription, uncertainties associated with this ambiguity are evaluated by varying $\bar{\Lambda}$ by a factor 2, namely by taking $1/2 \leq \bar{\Lambda}/(2\mu) \leq 2$.

Below, I review how the effect of the phase quenching appears in the perturbative expansion as described by Moore and Gorda in \cite{Moore:2023glb}.
One can construct a Feynman rule for each determinant in the square root in the phase-quenched fermion determinant~\eqref{eq:pqdet} in analogy to the ordinary perturbation theory.
In the Feynman rules of the phase-quenched theory, an additional factor $1/2$ arises from the following relation
\begin{equation}
    \sqrt{\det \calD_f(\pm\mu_f)} = \exp\left(\frac12 \tr \ln \calD_f(\pm \mu_f) \right)\,.
    \label{eq:detquench}
\end{equation}
Therefore, the only difference between the perturbative expansion of the phase-quenched theory and the original theory is that the sign of the $\mu$ is reversed for half of the fermions.

The phase-quenched theory and the original theory have the same perturbative expansion up to $\calO(\alpha_s^2)$.
One can explicitly verify that by setting $\mu \to - \mu$ in Eq.~\eqref{eq:ppqcd};
it does not change the expression of $P$.

The effect of $\mu \to -\mu$ appears at $\calO(\alpha_s^3)$.
This can be described schematically by going into the Nambu-Gorkov basis
\begin{equation}
    \label{eq:nambu}
    \Psi =
    \begin{pmatrix}
        \psi \\ \psi_C
    \end{pmatrix}\,,
\end{equation}
where $\psi_C = C\bar{\psi}^\top$ is the charge-conjugate spinor and $C \equiv i \gamma^2 \gamma^0$ is the charge conjugation operator.
In this basis, free fermion propagators are
\begin{equation}
    S_0^{-1} \equiv
    \begin{pmatrix}
        [G_0^+]^{-1} & 0 \\
        0 & [G_0^-]^{-1}
    \end{pmatrix}
    \,.
\end{equation}
where $[G_0^\pm]^{-1}(X,Y) \equiv -i (i\gamma^\mu \partial_\mu \pm \mu \gamma^0) \delta^{(4)}(X-Y)$.
The quark-gluon coupling is modified as
\begin{equation}
    \bar{\psi} \gamma^\mu T_a \psi A_\mu^a = \frac12 \bar{\Psi} \Gamma^\mu_a \Psi A_\mu^a\,,
\end{equation}
where
\begin{equation}
    \label{eq:qg_nambu}
    \Gamma^\mu_a \equiv
    \begin{pmatrix}
        \gamma^\mu T_a & 0 \\
        0 & - \gamma^\mu T_a^\top
    \end{pmatrix}\,.
\end{equation}
The upper and lower component of $\Psi$ gives the equivalent description of the theory.
Reversing the sign of $\mu$ in the upper component of $\Psi$ gives the equivalent description in the lower component of $\Psi$ with the original value of $\mu$ but with the quark-gluon coupling in the conjugate representation.
Therefore, flipping the sign $\mu \to -\mu$ is equivalent to changing $T_a \to -T_a^\top (=- T_a^\ast)$ in the quark-gluon vertex while keeping the original value of $\mu$ in the propagator, and the effect of $\mu \to -\mu$ only appears in the color factor of diagrams.
The difference in the diagrammatic expansion appears only at $\calO(\alpha_s^3)$.
This is due to the difference in color factors with the fundamental and antifundamental representations~\cite{Moore:2023glb}
\begin{equation}
        \tr T_a T_b T_c \neq \tr \left[(-T_a^\top) (- T_b^\top) (- T_c^\top)\right]\,.
\end{equation}
This color factor appears in the diagram with three gluons attached to two fermion loops involving different flavors.

\section{Cooper pairing in QCD at large isospin density}
\label{sec:cooper}

In this section, I first review the Cooper pairing in the isospin QCD.
Then, I derive and solve the gap equation perturbatively.

\subsection{Overview}

At large $\muI > 0$\footnote{Note that I take $\muI > 0$, which is opposite to the choice in Ref.~\cite{Son:2000xc, *Son:2000by}}, $u$ and $\bar{d}$ quarks fill up the Fermi sphere with the radius of $\muI/2$ in the ground state.
The Cooper instability leads to that these $u$ and $\bar{d}$ quarks form Cooper pairs in the color-singlet, pseudoscalar, and ${}^1S_0$ channel:
\begin{equation}
    \label{eq:pion}
    \langle \bar{d}_\alpha \gamma^5 u_\beta \rangle \propto \delta_{\alpha \beta} \Delta \,,
\end{equation}
where the greek letters $\alpha, \beta$ are the color indices in the fundamental representation and $\Delta$ is the gap parameter.
Note that it has the same quantum number as $\pi^+$, and this pattern of pairing is favored from the QCD inequality~\cite{Son:2000xc, *Son:2000by}.

The gap $\Delta$ can be calculated in a similar setup as in the diquark condensation at nonzero $\muB$.
In the weak coupling expansion, $\Delta$ on the Fermi surface has the form
\begin{equation}
    \log \left(\frac{\Delta}{\mu}\right) = - \frac{b_{-1}}{g} - \bar{b}_0 \ln g - b_0  + \cdots\,,
\end{equation}
where I have truncated the perturbation series at $\calO(1)$ and $\calO(\ln g)$.

In QCD at nonzero $\muI$, the coefficient $b_{-1}$ is
\begin{equation}
    b_{-1} = \sqrt{\frac{6\Nc}{\Nc^2 - 1}} \pi^2= \frac{3\pi^2}{2}\,,
\end{equation}
as first pointed out in Refs.~\cite{Son:2000xc, *Son:2000by}.
Note that this value is $1/\sqrt{2}$ times smaller compared to that in QCD at nonzero $\muB$, which is $b_{-1} = \sqrt{6\Nc / (\Nc + 1)} \pi^2 = 3\pi^2/\sqrt{2}$ first pointed out in Ref.~\cite{Son:1998uk} (see also Refs.~\cite{Hong:1999fh, Schafer:1999jg, Pisarski:1999bf, Brown:1999aq, Hsu:1999mp,  Pisarski:1999tv,Brown:1999yd, Brown:2000eh}.
Therefore the magnitude of $\Delta$ is exponentially enhanced.
For later convenience, I rewrite the remaining terms as
\begin{equation}
    - \bar{b}_0 \ln g - b_0 = \ln \btilde - b_0'\,.
\end{equation}
The factor $\btilde$ arises from the gluon sector, in which magnetic and electric gluon exchange occurs at a large angle, so it is independent of the color structure of the condensate and whether the chemical potential being $\muB$ or $\muI$.
It reads
\begin{equation}
    \label{eq:btilde}
    \btilde = 512 \pi^4 g^{-5}\,.
\end{equation}
The factor $b_0'$ arises from the wave-function renormalization.
In this work, I compute this factor for the first time at nonzero $\muI$, and I find
\begin{equation}
    b_0' = \frac{\pi^2 + 4}{16}\,,
\end{equation}
and the numerical value is $e^{-{b_0'}} \simeq 0.420$.
At nonzero $\muB$, it was calculated in Refs.~\cite{Brown:1999aq, Wang:2001aq}, and they found
\begin{equation}
    b_0' = \frac1{16} (\pi^2+ 4) (\Nc - 1) = \frac{\pi^2 + 4}{8}\,,
\end{equation}
with the numerical value $e^{-{b}_{0}'} \simeq 0.177$.

Summarizing these results, the superfluid gap $\Delta$ is concisely summarized as
\begin{equation}
    \label{eq:DeltaI}
    \Delta = \btilde \mu \exp\left(- \frac{\pi^2 + 4}{16}\right) \exp\left(- \frac{3\pi^2}{2 g}\right) \,.
\end{equation}
The gap function in QCD at $\muI \neq 0$ is exponentially enhanced compared to the color-superconducting gap in QCD at $\muB \neq 0$.
This is because the attraction arising from the one-gluon exchange is stronger in the color singlet $q\bar{q}$ channel compared to that in the color antitriplet $qq$ channel~\cite{Son:2000xc, *Son:2000by}.

In the following two subsections, I will derive and solve the gap equation for $\Delta$, and clarify the difference from the diquark condensation at nonzero $\muB$.

\subsection{Gap equation}

In this subsection, I derive the gap equation from the perturbation theory.
The gap equation is very similar to that of the two-flavor color superconductor, so I will follow the formalism developed in the context of color superconductivity, in which the Nambu-Gorkov basis is employed~\eqref{eq:nambu}.
The major difference is that the Cooper pairing occurs in $q\bar{q}$-channel, not in the diquark channel.
Consequently, the Nambu-Gorkov basis~\eqref{eq:nambu} is replaced with the isospin basis~\cite{Cohen:2015soa}, namely,
\begin{equation}
    \label{eq:isospinbasis}
    \Psi =
    \begin{pmatrix}
        \psi \\ \psi_C
    \end{pmatrix}\,
\to \,
    \Psi = 
    \begin{pmatrix}
        u \\ d
    \end{pmatrix}\,.
\end{equation}
I write down the gap equation in the isospin basis.
In this basis, the free quark propagators are
\begin{equation}
    \label{eq:S0}
    S_0^{-1} \equiv
    \begin{pmatrix}
        [G_0^+]^{-1} & 0 \\
        0 & [G_0^-]^{-1}
    \end{pmatrix}
    \,,
\end{equation}
The inverse free propagator is expressed in the momentum space as
\begin{equation}
    \label{eq:G0pm}
    [G_0^\pm]^{-1}  = \sum_{e = \pm} [k_0 \pm (\mu - ek)] \gamma^0 \Lambda_{\bk}^{(\pm e)}\,,
\end{equation}
where $\mu = \muI / 2$ is the quark chemical potential, and $\Lambda_{\bk}^{(e)}$ is the energy projector
\begin{equation}
    \Lambda_{\bk}^{(e)} \equiv \frac{1 + e \gamma^0 \bgamma \cdot \bkhat}2\,.
\end{equation}
The quark-gluon coupling $\bar{\Psi} \Gamma^\mu_a \Psi A_\mu^a$ in this basis is characterized by the following matrix
\begin{equation}
    \Gamma^\mu_a \equiv
    \begin{pmatrix}
        \gamma^\mu T_a & 0 \\
        0 & \gamma^\mu T_a
    \end{pmatrix}\,.
\end{equation}
Notice the difference between the quark-gluon vertex in the isospin basis~\eqref{eq:isospinbasis} and the Nambu-Gorkov basis~\eqref{eq:qg_nambu};  in the latter case, the lower component of $\Gamma^\mu_a$ is in the antifundamental representation $-T_a^\top$.

The quark part of the two-particle-irreducible (2PI) action $\Gamma$ can be written as the following functional of the full quark propagator $S$~\cite{Luttinger:1960ua, Baym:1962sx, deDominicis:1964zz, Cornwall:1974vz}
\begin{align}
    \label{eq:2PI}
    \Gamma[S] &= \tr \ln S^{-1} + \tr(S_0^{-1} S - 1) + \Gamma_2[D,S]\,,
\end{align}
where $\Gamma_2$ is the sum of the 2PI skeleton diagrams, and it also depends on the full gluon propagator $D$.
Notice that this expression does not include a factor $1/2$ as in Eq.~(14) of Ref.~\cite{Alford:2007xm}, which is required to cancel the double counting in the Nambu-Gorkov basis.
The ground state is given by the stationary point of $\Gamma$.
From the stationarity condition $\delta \Gamma[S] / \delta S = 0$, I obtain Schwinger-Dyson equation
\begin{equation}
    \label{eq:SD}
    S^{-1} = S_0^{-1} + \Sigma\,,
\end{equation}
where $\Sigma$ is the quark self-energy and defined by the functional derivative of $\Gamma_2$ at the stationary point
\begin{align}
    \label{eq:self}
    \Sigma \equiv \frac{\delta \Gamma_2}{\delta S}\,.
\end{align}

Here, I follow the conventional approximation for $\Gamma_2$ in which one truncates the infinite sum of the 2PI skeleton diagrams up to the two-loop order;
this two-loop approximation for $\Gamma_2$ corresponds to a one-loop approximation for $\Sigma$.
Then, the gap equation~\eqref{eq:self} becomes
\begin{equation}
    \label{eq:gap}
    \Sigma(K) = - g^2 \SumInt_{\!\! Q} \Gamma_a^\mu S(Q) \Gamma_b^\nu D_{\mu\nu}^{ab}(K-Q)\,,
\end{equation}
where $\SumInt_Q \equiv T\sum_{\omega_n} \int \frac{d^3\bq}{(2\pi)^3}$ denote the sum over the Matsubara modes and integration in the momentum space.
I write the quark self-energy in the isospin basis as
\begin{equation}
    \Sigma \equiv
    \begin{pmatrix}
        \Sigma^{+} & \Phi^{-} \\
        \Phi^{+} & \Sigma^{-}
    \end{pmatrix}\,,
\end{equation}
and the quark full propagator as
\begin{equation}
    \label{eq:S}
    S \equiv
    \begin{pmatrix}
        G^{+} & F^{-} \\
        F^{+} & G^{-}
    \end{pmatrix}\,.
\end{equation}
The off-diagonal elements of the self-energy are related via $\Phi^- = \gamma^0 (\Phi^+)^\dagger \gamma^0$.
Through the Schwinger-Dyson equation \eqref{eq:SD}, one can express the diagonal and anomalous propagators in terms of the free propagator and the self-energy as
\begin{align}
    \label{eq:fullG}
    G^{\pm} &= \left[ [G_0^{\pm}]^{-1} + \Sigma^{\pm} - \Phi^{\mp} \left([G_0^{\mp}]^{-1} + \Sigma^{\mp}\right)^{-1} \Phi^{\pm}\right]^{-1}\,,\\
    F^{\pm} &= \left([G_0^{\mp}]^{-1} + \Sigma^{\mp}\right)^{-1} \Phi^{\pm} G^{\pm}\,.  
    \label{eq:fullF}
\end{align}

The diagonal elements of the quark self-energy are calculated as~\cite{Manuel:2000nh, Brown:1999yd}
\begin{equation}
    \Sigma^{\pm} \simeq \gbar^2 \ln \left(\frac{M^2}{k_0^2}\right) k_0 \gamma^0 \Lambda_{\bk}^{\pm}\,.
\end{equation}
where $\gbar \equiv g / (3\sqrt{2} \pi)$ and $M^2 \equiv (3\pi/4) m_g^2$ with the gluon mass being $m_g^2 \equiv \Nf g^2 \mu^2 / (6\pi^2) $.

For the off-diagonal part of the quark self-energy, I use the following ansatz for the gap matrix given the pairing pattern in Eq.~\eqref{eq:pion}
\begin{equation}
    \label{eq:Phipm}
    \Phi^{\pm}(K) = \pm \Delta(K) \gamma^5 \calM \,,
\end{equation}
and $\calM$ is the matrix in the color space
\begin{equation}
    \calM_{\alpha\beta} = \delta_{\alpha\beta}\,.
\end{equation}

By substituting all these equations in Eqs.~\eqref{eq:fullG} and \eqref{eq:fullF}, I obtain
\begin{align}
    \label{eq:Gpm}
    G^\pm &= \left([G_0^{\mp}]^{-1} + \Sigma^{\mp}\right)
    \sum_{e} \frac{\calM \Lambda_{\bk}^{(\mp e)}}{[k_0 / Z^{(e)}(k_0)]^2  - [\epsilon_{k}^{(e)}]^2}\,,\\
    F^{\pm} &= \pm \Delta \gamma^5 \calM \sum_{e} \frac{\Lambda_{\bk}^{(\mp e)} }{[k_0 / Z^{(e)}(k_0)]^2  - [\epsilon_{k}^{(e)}]^2}\,,
    \label{eq:Fpm}
\end{align}
where the wave-function renormalization is defined as
\begin{align}
    Z^{(+)}(k_0) \equiv \left[1 + \gbar^2 \ln \left(\frac{M^2}{k_0^2}\right) \right]^{-1}\,,
\end{align}
and $Z^{(-)}(k_0) = 1$.
The quasiparticle energy is defined as
\begin{equation}
    \epsilon_{k}^{(e)} \equiv \sqrt{(ek - \mu)^2 + |\Delta^{(e)}|^2}\,.
\end{equation}

Now, I consider the (2, 1)-component of the gap equation~\eqref{eq:gap}
\begin{equation}
    \label{eq:gapeqphi}
    \Phi^+ (K) = - g^2 \SumInt_{\!\! Q} \gamma^\mu T_a F^{+}(Q) \gamma^\nu T_b D_{\mu\nu}^{ab}(K-Q)\,,
\end{equation}
where $D_{\mu\nu}^{ab}$ is the gluon propagator.
The gluon propagator is assumed to have the color structure $D^{ab} \propto \delta^{ab}$, where the roman indices $a, b, \ldots$ are the color indices in the adjoint representation.
By substituting $\Phi^+$~\eqref{eq:Phipm} and $F^+$~\eqref{eq:Fpm} in the equation above, multiplying $\calM^\dagger \gamma^5 \Lambda_{\bk}^{(+)}$ to both hand sides of the equation, and taking the trace, I obtain
\begin{align}
    \Delta(K) = &- C_F g^2 \SumInt_{\!\! Q} \frac{\Delta(Q)}{[q_0 / Z^{(+)}(q_0)]^2 - [\epsilon_q^{(+)}]^2} \notag \\
    &\times \frac{\tr \left[\gamma^\mu \gamma^5 \Lambda_{\bq}^{(-)} \gamma^\nu \gamma^5 \Lambda_{\bk}^{(+)}\right]}{\tr \Lambda_{\bk}^{(+)}} D_{\mu\nu}(K-Q)\,.
\end{align}
I neglected the antiparticle contribution.
Hereafter, I suppress the superscript $(+)$ as there is only a quasiparticle contribution.
The color factor $C_F$ appearing in front is
\begin{equation}
    \label{eq:colorI}
    C_F=\frac{\tr(T_a \calM T_a \calM^\dagger)}{\tr(\calM \calM^\dagger)} = \frac{\Nc^2 - 1}{2\Nc} = \frac{4}{3}\,.
\end{equation}
This color factor is associated with the $\boldsymbol{1}$ channel, which is the most attractive among the available color channels in the one-gluon exchange interaction between $q\bar{q}$, and the available color channels are given by the decomposition $\boldsymbol{3} \otimes \boldsymbol{\bar{3}} = \boldsymbol{1} \oplus \boldsymbol{8}$.
Note that this is twice as large as that in the 2SC phase of the two-flavor color superconductor at nonzero $\muB$, in which the color structure of the gap is antisymmetric $\calM_{\alpha\beta} = \epsilon_{\alpha\beta 3}$ and the color factor is
\begin{equation}
    \label{eq:colorB}
    \frac{\tr[(-T_a^T) \calM (\calM \calM^{\dagger}) T_a \calM^\dagger]}{\tr(\calM \calM^\dagger)} = \frac{\Nc + 1}{2\Nc} = \frac{2}{3}\,.
\end{equation}
This color factor is associated with the $\boldsymbol{\bar{3}}$ channel, which is the most attractive among the available color channels in the one-gluon exchange interaction between $qq$, and the available color channels are given by the decomposition $\boldsymbol{3} \otimes \boldsymbol{3} = \boldsymbol{\bar{3}} \oplus \boldsymbol{6}$.

After splitting the gluon propagator into the longitudinal and transverse component, taking the Matsubara sum, which is the same procedure as in the nonzero-$\muB$ case~\cite{Pisarski:1999tv}, the gap equation becomes
\begin{align}
    \Delta_k \simeq 2\gbar^2 \int_0^\delta \frac{d (q - \mu)}{\etilde_q} \bigg[&Z^2(\etilde_q) \tanh \left(\frac{\etilde_q}{2T}\right) \notag \\
    & \times \frac12 \ln \left(\frac{\btilde^2 \mu^2}{|\etilde_q^2 - \etilde_k^2|}\right) \Delta_q \bigg]\,,
    \label{eq:gapeq}
\end{align}
where I denote $\etilde_q \equiv Z(\epsilon_q) \epsilon_q$, and $\Delta_k \equiv \Delta(\etilde_k, \bk)$.
The factor $\btilde$ is defined in Eq.~\eqref{eq:btilde}.
Note that an additional prefactor $2 = \Nc -1$ arises in the isospin QCD by replacing the color factor \eqref{eq:colorB} in the baryonic QCD by \eqref{eq:colorI}.
I elaborate on the solution to this equation in the next subsection.

\subsection{Solution of the gap equation}
\label{sec:sol}
In this subsection, I clarify how the additional factor two in Eq.~\eqref{eq:gapeq} modifies the solution of the gap equation by explicitly solving it.
I follow the calculation at nonzero $\muB$ presented in Refs.~\cite{Pisarski:1999tv, Wang:2001aq} (the same results can be derived using different formalisms as in Refs.~\cite{Brown:1999aq, Brown:1999yd, Brown:2000eh, Schafer:2003jn}).
In the equations below, all the modifications arising at nonzero $\muI$ are underlined.
Namely, if all the underlined coefficients are discarded, one recovers the calculation at nonzero $\muB$ in Ref.~\cite{Wang:2001aq}.

I solve the gap equation~\eqref{eq:gapeq} at zero temperature to obtain the gap function at the Fermi surface $\Delta_\ast \equiv \Delta_{q=\mu}$.
The thermal factor becomes $\tanh[\etilde_q/(2T)] =1$.
As for the dressed energy $\etilde_q$ in logarithms, I approximate as $\ln(\btilde \mu / |\etilde_q^2 - \etilde_k^2|) \simeq \ln(\btilde \mu / |\epsilon_q^2 - \epsilon_k^2|)$ and $Z(\etilde_q) \simeq Z(\epsilon_q)$, which are valid at the leading order.
Then the gap equation becomes
\begin{equation}
    \Delta_k \simeq \changed{2}\gbar^2 \int_0^\delta \frac{d (q - \mu)}{\epsilon_q} Z(\epsilon_q) \frac12 \ln \left(\frac{\btilde^2 \mu^2}{|\epsilon_q^2 - \epsilon_k^2|}\right) \Delta_q\,,
\end{equation}
In Ref.~\cite{Son:1998uk}, Son observed that the logarithm can be replaced by $\max\{\ln(\btilde \mu / \epsilon_k), \ln(\btilde \mu / \epsilon_q)\}$ at this order,
so I make a further approximation
\begin{align}
    &\frac12 \ln \left(\frac{\btilde^2 \mu^2}{|\epsilon_q^2 - \epsilon_k^2|}\right) \notag \\
    &\simeq
    \ln\left(\frac{\btilde \mu}{\epsilon_q}\right) \theta(q-k) + \ln\left(\frac{\btilde \mu}{\epsilon_k}\right) \theta(k-q)\,,
\end{align}
and introduce the variables~\cite{Pisarski:1999tv, Wang:2001aq}
\begin{equation}
\begin{split}
    x &\equiv \gbar \ln\left(\frac{2\btilde \mu}{k - \mu + \epsilon_k}\right)\,,\\
    y &\equiv \gbar \ln\left(\frac{2\btilde \mu}{q - \mu + \epsilon_q}\right)\,,\\
    x_\ast &\equiv \gbar \ln\left(\frac{2\btilde \mu}{\Delta_\ast}\right)\,,\\
    x_0 &\equiv \gbar \ln\left(\frac{\btilde \mu}{\delta}\right)\,.
\end{split}
\end{equation}
Since $\Delta_\ast \sim \mu \exp(- 1 / \gbar)$, these variables scale in the $\gbar$ expansion as
\begin{equation}
\label{eq:scaling}
\begin{split}
    x, y &\sim
    \begin{cases}
        \calO(1) & \text{(close to the Fermi surface)}\,,\\
        \calO(\gbar) & \text{(away from the Fermi surface)}\,,
    \end{cases}\\
    x_\ast &\sim \calO(\gbar)\,,\\
    x_0 &\sim \calO(1)\,.
\end{split}
\end{equation}
The gap equation up to $\calO(\gbar)$ is written in terms of these variables as
\begin{align}
    \Delta(x) &\simeq \changed{2}x \int_{x}^{x_\ast} dy (1 - 2\gbar y) \Delta(y) \notag \\
    &\quad + \changed{2}\int_{x_0}^x dy y (1 - 2\gbar y) \Delta(y)\,. \label{eq:delta}
\end{align}

To determine the functional form of $\Delta(x)$, I take the second derivative of the gap equation~\eqref{eq:delta} to convert the integral equation into the differential equation
\begin{equation}
    \Delta''(x)
    \simeq - \changed{2} (1 - 2\gbar x) \Delta(x)\,,
    \label{eq:dddelta}
\end{equation}
where the $'$ symbol denotes the derivative with respect to the argument of the function.
By changing the independent variable from $x$ to $z \equiv - \changed{2^{1/3}} (2\gbar)^{-2/3} (1 - 2\gbar x)$, the equation~\eqref{eq:dddelta} becomes the Airy equation
\begin{equation}
    \label{eq:airy}
    \Delta''(z) - z \Delta(z) = 0\,.
\end{equation}
The solution to this equation is given by a linear combination of the Airy functions $\Ai$ and $\Bi$.
With arbitrary constants $C_1$ and $C_2$, $\Delta(z)$ is expressed as
\begin{equation}
    \Delta(z) = C_1 \Ai(z) + C_2 \Bi(z)\,.
\end{equation}
For later convenience, the Airy functions are decomposed into the phase and modulus as
\begin{align}
    \Ai(x) &= M(|z|) \cos\theta(|z|)\,, & \Bi(z) &= M(|z|) \sin \theta(|z|)\,, \notag \\
    M(|z|) &= \sqrt{\Ai^2(z) + \Bi^2(z)}\,, & \theta(|z|) &= \arctan\left(\frac{\Ai(z)}{\Bi(z)}\right)\,.
\end{align}
and also for the derivative of the Airy functions, I decompose as
\begin{align}
    \Ai'(x) &= N(|z|) \cos\varphi(|z|)\,, &\Bi'(z) &= N(|z|) \sin \varphi(|z|)\,, \notag \\
    N(|z|) &= \sqrt{\Ai'^2(z) + \Bi'^2(z)}\,, & \varphi(|z|) &= \arctan\left(\frac{\Ai'(z)}{\Bi'(z)}\right)\,,
\end{align}
The coefficients $C_1$ and $C_2$ are fixed by the boundary conditions $\Delta(z_\ast) = \Delta_\ast$ and $\Delta'(z_\ast) = 0$, where $z_\ast = - \changed{2^{1/3}} (2\gbar)^{-2/3} (1 - 2\gbar x_\ast)$.
The solution $\Delta(z)$ and its derivative are
\begin{align}
    \label{eq:deltasol}
    \Delta(z) &= \Delta_\ast \frac{M(|z|)}{M(|z_\ast|)} \frac{\sin\left[\varphi(|z_\ast|) - \theta(|z|)\right]}{\sin\left[\varphi(|z_\ast|) - \theta(|z_\ast|)\right]}\,,\\
    \Delta'(z) &= \Delta_\ast \frac{N(|z|)}{M(|z_\ast|)} \frac{\sin\left[\varphi(|z_\ast|) - \varphi(|z|)\right]}{\sin\left[\varphi(|z_\ast|) - \theta(|z_\ast|)\right]}\,.
    \label{eq:ddeltasol}
\end{align}

Now, I have determined the functional form of $\Delta(z)$.
For the remaining part, I need to evaluate the undetermined constant $\Delta_\ast$.
To this end, I set $x = x_\ast$ in Eq.~\eqref{eq:delta} and change the integration variable as $y \to w \equiv - \changed{2^{1/3}} (2\gbar)^{-2/3}(1 - 2\gbar y)$, then I arrive at
\begin{equation}
    \Delta(z_\ast) = \int_{z_\ast}^{z_0} dw\left[w + \changed{2^{1/3}} (2\gbar)^{-2/3}\right]w\Delta(w)\,.
\end{equation}
The function $w \Delta(w)$ in the integral can be replaced with $\Delta''(w)$ using the Airy equation~\eqref{eq:airy}.
Then, the integration by part gives the following relation
\begin{equation}
    \left[z_0 + \changed{2^{1/3}}(2\gbar)^{-2/3}\right] \Delta'(z_0) - \Delta(z_0) = 0\,,
\end{equation}
where $z_0 = - \changed{2^{1/3}} (2\gbar)^{-2/3} (1 - 2\gbar x_0)$.
By substituting Eqs.~\eqref{eq:deltasol} and \eqref{eq:ddeltasol} into the above equation, I obtain
\begin{align}
    \changed{2^{1/3}}(2\gbar)^{1/3} x_0 \sin\left[\varphi(|z_\ast|) - \varphi(|z_0|)\right]& \notag\\
    - \frac{M(|z_0|)}{N(|z_0|)}\cos\left[\varphi(|z_\ast|) - \theta(|z_0|) - \frac{\pi}2\right] &= 0\,.
\end{align}
I expand this equation up to the next-to-leading order in terms of $\gbar$.
In the above equation, I use the following asymptotic formulae, which are valid at weak coupling, $|z| \sim \gbar^{-2/3} \gg 1$
\begin{align}
    \varphi(|z|) &\simeq \frac{3\pi}{4} - \frac23 |z|^{3/2} - \frac{7}{48}|z|^{-3/2} + \calO(|z|^{-9/2}) \notag \\
    & \simeq - \frac{\changed{\sqrt2}}{3\gbar} + \frac{3\pi}{4} + \changed{\sqrt{2}} x - \gbar \left(\frac{\changed{\sqrt{2}}x^2}{2} + \frac{7}{24 \changed{\sqrt{2}}}\right) + \calO(\gbar^2)\,,\\
    \theta(|z|) &\simeq \frac{\pi}{4} - \frac23 |z|^{3/2} + \frac5{48} |z|^{-3/2} + \calO(|z|^{-9/2}) \notag \\
    & \simeq - \frac{\changed{\sqrt2}}{3\gbar} + \frac{\pi}{4} + \changed{\sqrt{2}} x - \gbar \left(\frac{\changed{\sqrt{2}}x^2}{2} - \frac{5}{24 \changed{\sqrt{2}}}\right) + \calO(\gbar^2)\,,\\
    \frac{M(|z|)}{N(|z|)} &\simeq |z|^{-1/2} + \calO(|z|^{-7/2}) \notag\\
    &\simeq \changed{2^{-1/6}} (2\gbar)^{1/3} \left[ 1 + \gbar x + \calO(\gbar^2) \right]\,,
\end{align}
and I get
\begin{align}
    \changed{2^{1/3}} (2\gbar) \bigg\{ &x_0 \sin(\changed{\sqrt{2}}x_\ast) - \frac1{\changed{\sqrt{2}}}\bigg[\cos(\changed{\sqrt{2}}x_\ast) \notag \\
    &+ \frac{\gbar}{2} \left(\changed{\sqrt{2}} x_\ast^2+ \frac{2\changed{\sqrt{2}} x_0}{\gbar} + \frac1{\changed{\sqrt2}}\right) \sin(\changed{\sqrt{2}}x_\ast) \bigg]\notag \\
    &+ \calO(\gbar^2) \bigg\} =0\,.
\end{align}
I note that I used the fact that $x_0$ is one order higher than $x_\ast$ based on the scaling behavior~\eqref{eq:scaling}.
Also, $x_0$ is an arbitrary scale and far from the Fermi surface, so it is expected that $x_0$ dependence cancels in the final expression.
Indeed, the $x_0$-dependence terms cancel in the equation above, and results become independent of $x_0$ up to $\calO(\gbar)$ in the perturbative expansion.
From this relation,
\begin{equation}
    \changed{\sqrt{2}} x_\ast \simeq \arctan\left[- \frac{2}{\gbar(1 / \changed{\sqrt{2}} + \changed{\sqrt{2}} x_\ast^2)}\right]\,,
\end{equation}
and its expansion owing to the relation $\arctan(-1/x) \simeq \pi/2 + x + \calO(x^3)$ for $|x|\ll 1$ yields
\begin{equation}
    \changed{\sqrt{2}} x_\ast \simeq \frac{\pi}{2}  + \frac{\gbar}{2} \left(\changed{\sqrt{2}} x_\ast^2 + \frac{1}{\changed{\sqrt{2}}}\right)\,.
\end{equation}
By solving this quadratic equation up to $\calO(\gbar)$, and using the relation $x_\ast \equiv \gbar \ln(2\btilde \mu / \Delta_\ast)$, I finally arrive at
\begin{equation}
    \Delta_\ast \simeq 2\btilde \mu \exp\left[- \frac{\pi}{2\changed{\sqrt{2}}\gbar} - \frac{1}{\changed{2}} \left(\frac{1}{2} + \frac{\pi^2}{8}\right) + \calO(\gbar^2)\right]\,.
\end{equation}

\section{Pairing gap contribution to the equation of state}
\label{sec:cond}

In this section, I calculate the condensation energy of the BCS state up to $\calO(g)$.
And, I numerically evaluate the magnitude of such a correction and compare it with the lattice QCD data.

\subsection{Calculation of the condensation energy}

The physical pressure is obtained as the value of $\Gamma[S]$ at its extremum.
The stationarity condition implies the relation \eqref{eq:self}, and formally this relation can be rewritten as $\Gamma_2[S] = \frac12 \tr(\Sigma S)$.
By substituting this relation into the expression of $\Gamma[S]$~\eqref{eq:2PI} with the use of the Schwinger-Dyson equation, I obtain the pressure
\begin{equation}
    P = \tr \ln S^{-1} - \frac12 \tr (1 - S_0^{-1} S)\,.
\end{equation}
By substituting the bare quark propagator \eqref{eq:S0} and \eqref{eq:G0pm}, and the full quark propagator \eqref{eq:S}, \eqref{eq:Gpm}, and \eqref{eq:Fpm} into the above expression, and then by taking the Matsubara sum, 
 the pressure is obtained (the detailed derivation is presented in Sec. 2.4 in Ref.~\cite{Schmitt:2004hg})
\begin{align}
    P(\Delta) &= 2\Nc \sum_{e = \pm} \int \!\! \frac{d^3 \boldsymbol{k}}{(2\pi)^3} \left[ \etilde_k^{(e)} + 2T \ln\left(1+ e^{-\etilde_k^{(e)}/T}\right)\right. \notag \\
    & \qquad \qquad - Z^2 (\etilde_k^{(e)})\left. \frac{|\Delta|^2}{2\etilde_k^{(e)}} \tanh \left(\frac{\etilde_k^{(e)}}{2T}\right)\right]\,.
\end{align}
At zero temperature, the pressure becomes
\begin{align}
    P(\Delta) = 2\Nc \int \!\! \frac{d^3 \boldsymbol{k}}{(2\pi)^3} \left( \etilde_k - Z^2(\etilde_k)\frac{\Delta^2}{2\etilde_k} \right)\,,
\end{align}
where I suppressed the superscript ${(+)}$ of the dressed quasiparticle energy $\etilde_k$ since only the $(+)$ component has the contribution from the pairing gap $\Delta$;  the $(-)$ component is the same as in the unpaired vacuum.
Henceforth, the logarithm can be approximated as $Z(\etilde_k) \simeq Z(\epsilon_k)$ so that
\begin{align}
    P(\Delta) \simeq 2\Nc \int \!\! \frac{d^3 \boldsymbol{k}}{(2\pi)^3} Z(\epsilon_k) \left( \epsilon_k - \frac{\Delta^2}{2\epsilon_k} \right)\,.
\end{align}

The condensation energy is defined as
\begin{equation}
    \delta P \equiv P(\Delta) - P(\Delta=0)\,.
\end{equation}
Now, I limit the integral around the Fermi surface where $-\delta \leq k-\mu  \leq \delta$, and neglect the momentum dependence of the gap $\Delta$.
Around the Fermi surface, the density of states is $\mu^2 / (2\pi^2)$, and I use the integration parity to limit the range of integration to $[0, \delta]$.
Then the condensation energy is reduced to
\begin{align}
    \delta P
    = 2\Nc \frac{\mu^2}{\pi^2} \int_0^{\delta} \! \! d (k-\mu) \, \bigg[ & Z(\epsilon_k) \left( \epsilon_k - \frac{\Delta^2}{2\epsilon_k}\right)\notag \\
    &-Z(\epsilon_k^{(0)}) \epsilon_k^{(0)} \bigg]\,,
\end{align}
where $\epsilon_k^{(0)} = |k - \mu|$ is the quasiparticle energy in the unpaired vacuum.
By expanding $Z(\epsilon) \simeq 1 + 2\gbar^2 \ln[\epsilon_k / (\btilde \mu)] + \calO(\gbar^2)$
and defining $\xi \equiv k - \mu$
\begin{align}
    \delta P= 2\Nc \frac{\mu^2}{\pi^2} \int_0^{\delta} d \xi \,&\Bigg\{ \left[1 + 2 \gbar^2  \ln \left(\frac{\epsilon_\xi}{\btilde \mu}\right)\right] \left( \epsilon_\xi - \frac{\Delta^2}{2\epsilon_\xi}\right)\notag\\
    & \quad - \left[1 + 2\gbar^2 \ln \Big(\frac{\xi}{\btilde\mu}\Big)\right] \xi \Bigg\} \,.
\end{align}

The leading-order contribution to $\delta P$ in the expansion in terms of the coupling constant $g$ is
\begin{align}
    \delta P_{\rm LO} &\equiv 2\Nc \frac{\mu^2}{\pi^2} \int_0^{\delta} d \xi \, \left( \epsilon_\xi - \frac{\Delta^2}{2\epsilon_\xi} - \xi\right)\notag\\
    &\simeq \Nc \frac{\mu^2 \Delta^2}{2 \pi^2} \,,
\end{align}
where in the last line, I only kept the term that is leading in the expansion in terms of $\Delta /\delta$.

The next-to-leading-order contribution to $\delta P$ of $\calO(g)$ is
\begin{align}
    \delta P_{\rm NLO} &\equiv 4 \gbar^2 \Nc \frac{\mu^2}{\pi^2} \int_0^{\delta} d \xi \, \bigg[ \ln \left(\frac{\epsilon_\xi}{\btilde\mu} \right)\left( \epsilon_\xi - \frac{\Delta^2}{2\epsilon_\xi}\right) \notag\\
    &\qquad\qquad\qquad\qquad\quad - \ln \left(\frac{\xi}{\btilde\mu}\right) \xi \bigg] \,,\notag\\
    &\simeq \gbar^2 \Nc \frac{\mu^2 \Delta^2}{\pi^2} \bigg[\frac12
    + 2\ln \left(\frac{\delta}{\btilde\mu}\right) - \ln\left(\frac{\Delta}{2\btilde\mu}\right) \bigg]\,.
\end{align}
Again, in the last line, I kept the term that is leading in the expansion in terms of $\Delta / \delta$, which is found to be proportional to $\Delta^2$.
Among the terms in the square bracket in the last line, I only keep the last term $-\ln[\Delta/(2\btilde \mu)]$.
From the scaling in Eq.~\eqref{eq:scaling}, this is the only term at $\calO(1)$ and the other terms are $\calO(g)$. 
This is because of $\Delta \sim e^{-1/g}$, so $\ln \Delta \sim 1/g$ reduces the power of $g$ by one.
One can also think of this as absorbing the $\sim \ln \delta$ term into the definition of $\Delta^2$ in the prefactor since $\Delta$ has an implicit $\delta$-dependence neglected in the derivation of $\Delta$ above;  although it is not confirmed whether the $\delta$-dependence in $\Delta$ and the $\sim \ln\delta$-term match or not.
Therefore, the NLO contribution at $\calO(g)$ to $\delta P$ is
\begin{align}
    \delta P_{\rm NLO}&\simeq - \gbar^2 \Nc \frac{\mu^2 \Delta^2}{\pi^2} \ln\left(\frac{\Delta}{2\btilde\mu}\right) + \calO(\gbar^2)\,,\notag\\
    &= g \Nc \frac{\mu^2 \Delta^2}{12 \pi^2} + \calO(g^2)\,.
\end{align}

Summarizing the result, the condensation energy $\delta P$ up to $\calO(g)$ is
\begin{equation}
    \label{eq:pairing}
    \delta P = \frac{\Nc}{2\pi^2} \mu^2 \Delta^2 \left(1 + \frac{g}6 \right)\,.
\end{equation}
I will plot the numerical value of this term in the next subsection.

\subsection{Numerical results}
\begin{figure}
    \centering
    \includegraphics[width=0.95\columnwidth]{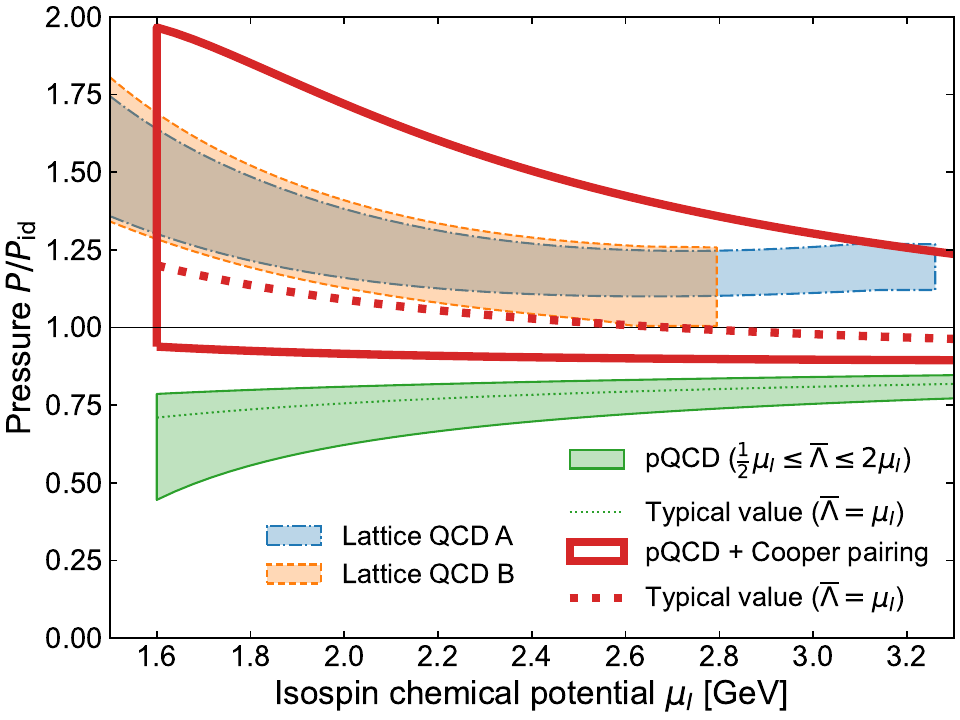}
    \caption{Comparison of the pQCD pressure with the lattice QCD data.  The bands for the pQCD results correspond to uncertainties arising from the ambiguity in the renormalization scale $\Lambdabar$, and dotted lines in the middle denote the common choice $\Lambdabar = \muI$. The pressure is normalized by the ideal-quark-gas value \eqref{eq:pid}.}
    \label{fig:ppid}
\end{figure}

In Fig.~\ref{fig:ppid}, I plot the isospin matter pressure normalized with the ideal gas value~\eqref{eq:pid}.
At $\mu \sim 1~\GeV$ (equivalently, $\muI \sim 2~\GeV$), it is expected that the pQCD works since the typical interaction scale in the system is large enough compared to the QCD scale $\Lambda_{\overline{\rm MS}}$.
Meanwhile, the lattice QCD calculation provides the EoS up to $\muI \sim 3~\GeV$~\cite{Abbott:2023coj}.
As one can see in Fig.~\ref{fig:ppid}, there is a discrepancy between these two calculations.

On the one hand, the pQCD calculation predicts $P/\Pid \simeq 1 - 2\alpha_s / \pi + \calO(\alpha_s^2) < 1$ and an increase in $P/ \Pid$ with increasing $\muI$ (the green band in Fig.~\ref{fig:ppid}).
For the pQCD calculation, I use the expression \eqref{eq:ppqcd}, and evaluate the scale variation uncertainty by taking $1/2 \leq \Lambdabar / \muI \leq 2$.
The lines in the green band from the bottom to the top correspond to the value $\Lambdabar / \muI = 1/2$, $1$, and $2$.

On the other hand, the lattice QCD data~\cite{Abbott:2023coj} dictates $P / \Pid > 1$ and $P/ \Pid$ decreases with increasing $\muI$ (the blue and orange bands in Fig.~\ref{fig:ppid}).
The blue and orange shaded bands marked with Lattice QCD A and Lattice QCD B are the results sampled from different ensembles;
the lattice geometry is $L^3 \times T = 48^3 \times 96$ and $64^3 \times 128$ for the ensemble A and B, respectively.

As explained in Sec.~\ref{sec:pert}, there can be a difference of $\calO(\alpha_s^3)$ between these two within the perturbation theory, however, the discrepancy is clearly larger than $\calO(\alpha_s^3)$.

Now, I add the condensation energy of the BCS state $\delta P$~\eqref{eq:pairing} to the pQCD pressure.
The red lines in the figure show the perturbative estimate with the Cooper pairing taken into account.
For the gap function, I use the expression in Eq.~\eqref{eq:DeltaI}.
The red lines from the bottom to the top correspond to the value $\Lambdabar / \muI = 2$, $1$, and $1/2$.
Note that this is in reverse order of the pQCD pressure without the pairing correction.
The smaller $\Lambdabar / \muI$ corresponds to the larger value of $\alpha_s$, and the gap is large for the larger value of $\alpha_s$.

One can see that the discrepancy is resolved by adding the condensation energy.
I stress that this result is without any fine-tuning, and the only ambiguity in the calculation is the choice of the renormalization scale $\Lambdabar$.

\begin{figure}
    \centering
    \includegraphics[width=0.95\columnwidth]{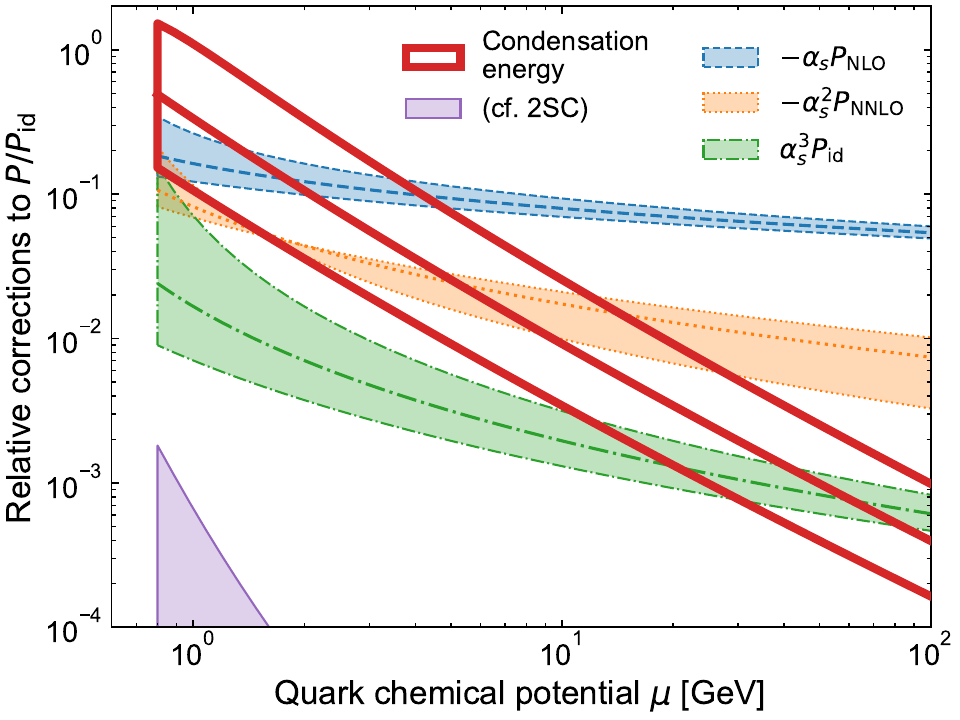}
    \caption{The magnitude of the perturbative and pairing corrections to $P$ relative to the ideal quark gas value $\Pid$.  The central line corresponds to the renormalization scale $\Lambdabar = 2\mu$, and the band around it represents $\Lambda=\mu$ and $4\mu$. Note that the horizontal axis is quark chemical potential $\mu$.}
    \label{fig:rel}
\end{figure}

In Fig.~\ref{fig:rel}, I compare the relative magnitudes of different contributions to the pressure.
Note that the $x$-axis is the quark chemical potential $\mu$, not the isospin chemical potential $\muI$.
I show the perturbative corrections to $P$ at each order, NLO~\eqref{eq:PNLO} and NNLO~\eqref{eq:PNNLO} relative to the ideal quark gas pressure $\Pid$~\eqref{eq:pid}.
I also plot $\calO(\alpha_s^3)$ contribution with $\alpha_s^3 \Pid$ mimicking the higher order corrections as the full contribution at $\calO(\alpha_s^3)$ is yet incomplete~\cite{Gorda:2023mkk}.
This is the order of magnitude expected for the difference between the phase-quenched theory and the original theory because this difference does not contain any logarithmically enhanced contribution $\sim \alpha_s \ln (\alpha_s)$ as reported in Ref.~\cite{Moore:2023glb}.

I plot the condensation energy~\eqref{eq:pairing} by the red lines.
The condensation energy contribution surpasses the dominant corrections in the pQCD at the NLO around $\mu \sim 1$-$5~\GeV$, depending on the renormalization scale.
Note that the scale variation uncertainty becomes larger for pQCD with the pairing contribution.

I also overlay the pairing gap contribution diquark gap in the 2SC phase in Fig.~\ref{fig:rel}.  The expression of the 2SC gap is
\begin{equation}
    \label{eq:Delta2SC}
    \Delta_{\rm 2SC} = \btilde \mu \exp\left(- \frac{\pi^2 + 4}{8}\right) \exp\left(- \frac{3\pi^2}{\sqrt{2} g}\right) \,.
\end{equation}
As expected, this contribution is exponentially suppressed and does not contribute to bulk thermodynamics.

\section{Quark-hadron crossover}
\label{sec:qhc}
The results presented above may also suggest the quark-hadron crossover from the BEC phase to the BCS phase.
At the equation level, one can see that the pion condensate in the BEC phase changes into a BCS condensate.

At low density, from the chiral perturbation, the leading contribution to the pressure is $\propto \muI^2$, and it reads~\cite{Son:2000xc, *Son:2000by, Adhikari:2019mdk}
\begin{equation}
    \label{eq:becterm}
    P_{I} \supset \frac{f_\pi^2}2 \muI^2\,.
\end{equation}
I call it the \emph{BEC term}.
From the one-loop correction in the chiral perturbation theory, the term $\propto \muI^4$ arises with a small prefactor $\sim 1/(4\pi)^{2}$~\cite{Adhikari:2019mdk}.
As the density becomes larger, the $\muI^4$-term becomes more relevant compared to the $\muI^2$-term.

From the calculations above, even at $\muI \simeq 2~\GeV$, I find there is a substantial contribution from the pairing in the BCS regime, which has the form
\begin{equation}
    \label{eq:bcsterm}
    P_{I} \supset \frac{\Nc \Delta^2}{2\pi^2} \left(1 + \frac{g}6 \right)\left(\frac{\muI}{2}\right)^2\,.
\end{equation}
I call it the \emph{BCS term}.
Although there is a $\mu$-dependence in $\Delta$, it varies slowly with increasing $\mu$, so this $\mu$-dependence is mild, and $\Delta$ can be regarded roughly as a constant.
Therefore, it has the same structure as the BEC term ~\eqref{eq:becterm}.
Namely, the BCS term can be rewritten as $\propto f_\Delta^2 \muI^2 / 2$ by defining the prefactor $f_{\Delta}$ as
\begin{equation}
    f_\Delta^2 \simeq \Nc \frac{\Delta^2}{4\pi^2} \left(1 + \frac{g}6 \right)\,.
\end{equation}
In the pQCD, there also arises a term with $\propto \muI^4$~\eqref{eq:pid}.

\begin{figure}
    \centering
    \includegraphics[width=0.95\columnwidth]{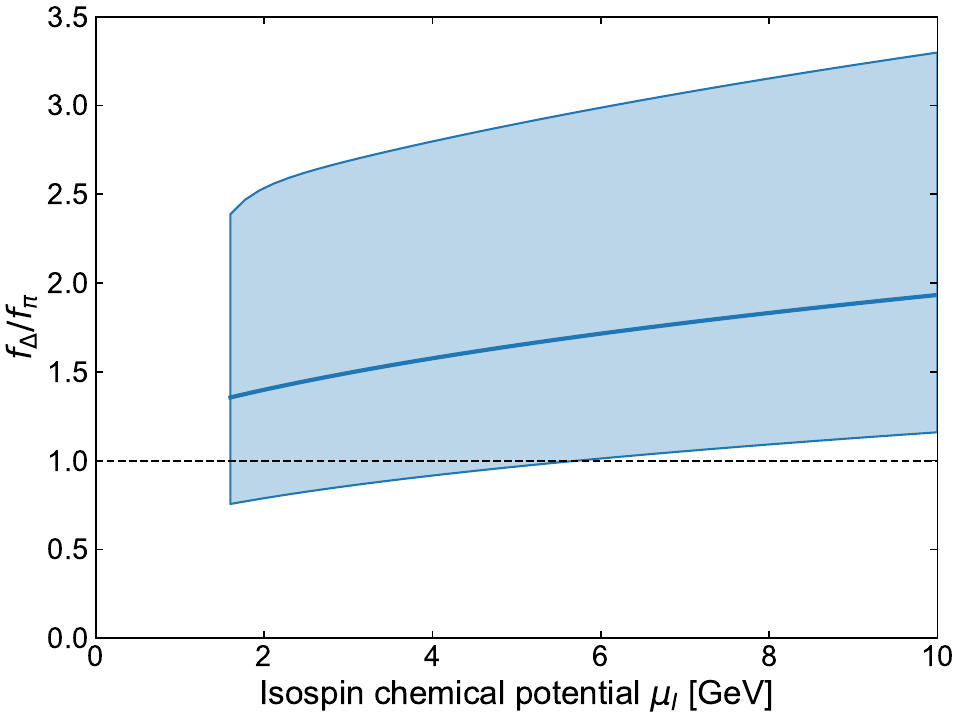}
    \caption{Relative magnitude of the prefactor of the BCS term compared to that of the BEC term.
    The vertical dashed line is the value of the pion decay constant in the vacuum, $f_\pi \simeq 93~\MeV$.
    The band shows the scale-variation uncertainty.}
    \label{fig:fpi}
\end{figure}
In Fig.~\ref{fig:fpi}, I show the relative magnitude of the prefactor of the BCS term, $f_\Delta$, to the pion decay constant $f_\pi$, which is the prefactor of the BEC term.
I find that the value of $f_\Delta$ stays close to $f_\pi$ and varies slowly with increasing $\muI$.
Note that the band in Fig.~\ref{fig:fpi} corresponds to the scale-variation uncertainty.

It may also be interesting to see the behavior of these terms in the large-$\Nc$ limit as discussed in Refs.~\cite{Son:2000xc, Son:2000by} (see also Refs.~\cite{Hanada:2011ju, Hanada:2012es}).
At small $\muI$, the BEC term \eqref{eq:becterm} scales as $\calO(\Nc)$ since $f_\pi^2 \sim \calO(\Nc)$.
At large $\muI$, the bulk thermodynamics is dominated by $\Pid = \Nc \Nf \mu^4 / (12\pi^2)$~\eqref{eq:pid}, which also scales as $\calO(\Nc)$.
Therefore, the BEC and BCS regime is continuous in the $\Nc$ scaling, which is similar to that at nonzero $\muB$~\cite{McLerran:2007qj} although the physics origin at small chemical potential is different;
in the nonzero-$\muI$ case, the $\Nc$ dependence arises from the chiral symmetry breaking through the Gell-Mann--Oakes--Renner relation, while in the nonzero-$\muB$ case, it arises from the interaction among baryons.

Now I turn to the large-$\Nc$ limit of the BCS term.
As I will explain shortly, it scales differently from the other terms in the pressure.
Unlike the BCS term in the color-superconducting phase, which is suppressed in the limit $\Nc \to \infty$ with fixed 't Hooft coupling $\lambda = g^2 \Nc$~\cite{Buchoff:2009za}, the BCS term at nonzero $\muI$ is nonvanishing in the large-$\Nc$ limit as this is a color singlet.
The naive estimate of the gap in the large-$\Nc$ by taking this limit in the expression \eqref{eq:DeltaI} gives
\begin{equation}
    \Delta \sim \mu \left(\frac{\Nc}{\lambda}\right)^{5/2} \exp\left(- \frac{\sqrt{6} \pi^2}{\sqrt{\lambda}}\right)\,.
\end{equation}
With this, the BCS term \eqref{eq:bcsterm} scales as $\calO(\Nc^6)$ although it is parametrically small compared to the bulk $\mu^4$-term for a small value of $\lambda$.
The prefactor $(\Nc / \lambda)^{5/2}$ in the above expression arises from the Debye screening and the Landau damping effects in the gluon exchange.
Therefore, the BCS term \eqref{eq:bcsterm} in the large-$\Nc$ limit scales differently from the pion BEC term \eqref{eq:becterm} as well as the bulk $\mu^4$-term.

However, these effects responsible for the large prefactor $(\Nc / \lambda)^{5/2}$ are suppressed by $\calO(\Nc^{-1/2})$ in the large-$\Nc$ limit, therefore the naive estimate above may be modified and gives the consistent $\Nc$-counting also for the BCS term.
I will justify this by the hand-waving estimate.
The gap equation takes the form~\cite{Alford:2007xm}
\begin{equation}
    \Delta \propto g^2 \frac{\Nc^2 - 1}{2\Nc}\int d\xi \frac{\Delta}{\sqrt{\xi^2 + \Delta^2}} d\theta \frac{\mu^2}{\theta \mu^2 + \delta^2}\,,
\end{equation}
where $\theta$ is the angle between the momenta $\boldsymbol{k}$ and $\boldsymbol{q}$ (see Eq.~\eqref{eq:gapeqphi} for definition) and $\delta$ is a cutoff scale for the collinear divergence.
At finite $\Nc$, $\delta$ arises from the Landau damping $\delta \sim (\Delta m_g^2)^{1/3}$.
Here, if we assume that the confinement persists at large $\Nc$, which is the fundamental assumption in Quarkyonic matter~\cite{McLerran:2007qj}, $\delta $ can be taken as the QCD scale $\Lambda_{\rm QCD}$, which is the characteristic scale for the confinement.
Then, the solution of the gap equation becomes $\Delta / \mu \sim \exp(- \Lambda_{\rm QCD}/\lambda)$ without the $\Nc^{5/2}$ factor in front.
Even though the parametric dependence of the gap on $\lambda$ is different, it still survives in the large-$\Nc$ limit.
This will give the same large-$\Nc$ scaling for the BCS term as the BEC term, so the quark-hadron crossover is implied from the large-$\Nc$ limit.

I also note that the gap parameter here wins over the gap of the chiral density wave, such as of the Deryagin, Grigoriev, and Rubakov (DGR) type~\cite{Deryagin:1992rw, Shuster:1999tn, Kojo:2009ha} (see also \cite{Kojo:2011cn}).
It is natural to expect so as the chiral density wave uses only the part of the phase space near the Fermi surface.
The detailed analysis will be reported elsewhere.

\section{Summary and discussion}
In this work, I studied QCD at nonzero isospin chemical potential $\muI \neq 0$ as a specific example of the phase-quenched theory.
I calculated the gap parameter and the condensation energy associated with it up to the next-to-leading order in the expansion in terms of the coupling constant $g$.
I found an exponential enhancement in this contribution, and the inclusion of this nonperturbative correction to the equation of state explains the discrepancy between the lattice QCD and naive pQCD results without any fine-tuning, as shown in Fig.~\ref{fig:ppid}.

This implies that when extracting the perturbative coefficients of $\calO(\alpha_s^4)$ from the phase-quenched lattice simulation, one needs to take into account the non-perturbative correction arising from the Cooper pairing.
In the physical sense, the effect of the phase quenching of the fermion determinant in the partition function is interpreted as an enhancement in the pairing gap (it is evident in Fig.~\ref{fig:rel}).
This correction cannot be treated within the perturbation theory explained in Sec.~\ref{sec:pert}.
Each fermion determinant in the square root of Eq.~\eqref{eq:pqdet} corresponds to quarks with positive and negative chemical potentials, and these quarks are treated separately in the perturbation theory.
However, in the actual lattice QCD calculation, there is a large contribution to the thermodynamics from the mixing between these quarks with positive and negative chemical potentials.

This can be exemplified clearly by relabeling the quarks with positive and negative chemical potentials in Eq.~\eqref{eq:pqdet} as $u$ and $d$ quarks, respectively.
Then, the Cooper pair condensation $\langle \bar{d} \gamma^5 u \rangle$ mixes $u$ and $d$ quarks, and one can have a diagram with $u$ and $d$ quarks running inside already at a one-loop level.
One can generalize the results in this paper from $\Nf=2$ to even $\Nf$  straightforwardly.  For odd $\Nf$, the generalization is more non-trivial, namely, the gap equation and its solution presented in Sec.~\ref{sec:cooper} should be modified to include an additional factor of 1/2 in Eq.~\eqref{eq:detquench} in the perturbation theory.
It would also be interesting to see the effect of flavor mixing of different origins, for example from the instanton-induced interaction (see, e.g.~\cite{Schafer:2002ty}).

To the best of my knowledge, this is the first cross-validation of the perturbative QCD at large density confronted with the lattice simulation.
This has several implications and impacts on QCD at nonzero baryon chemical potential, and possibly on neutron star physics.

I also verified that the calculation of the pairing gap as a solution of the gap equation derived in the perturbation theory is reliable in the regime where the perturbative expansion of the partition function is valid.
It is highly plausible that the evaluation of the pairing gap is reliable as well in QCD at nonzero \emph{baryon} chemical potential.
Posit that the weak-coupling calculation of the pairing gap is still valid around $\mu \sim 1~\GeV$, the pairing gap at nonzero baryon chemical potential is exponentially small compared to the isospin-QCD counterpart.
As a consequence, this fact implies that the color-superconducting gap does not affect the bulk properties such as the equation of state at least in the perturbative regime.
It further implies that the behavior of the trace anomaly introduced in Ref.~\cite{Fujimoto:2022ohj} is very different between QCD at nonzero $\muI$ and $\muB$.
Namely, the former case shows the large negative value for the trace anomaly as shown in Ref.~\cite{Abbott:2023coj} (see also Ref.~\cite{Chiba:2023ftg}), which is mainly caused by the pairing gap term in thermodynamics, while in the latter case, the trace anomaly can still be positive owing to the absence of the large pairing gap term as conjectured in Ref.~\cite{Fujimoto:2022ohj}.

As a future extension of this work, changing the number of colors $\Nc$ is an interesting direction, particularly taking $\Nc = 2$ and $\Nc \to \infty$.
In two-color QCD, one can use technology very similar to the present work to calculate the equation of state.
Two-color QCD at nonzero chemical potential is a theory free from sign problem, so one can confront the lattice data (to date, there are several lattice equation of state data available, e.g.~\cite{Hands:2006ve, Cotter:2012mb, Boz:2019enj, Begun:2022bxj, Iida:2022hyy}).
One may also expect the large diquark gap as well because the representations $\boldsymbol{2}$ and $\bar{\boldsymbol{2}}$ are equivalent in SU(2) due to the pseudoreality.
As I mentioned partially in the text, the expression for the pairing gap may be different in the large-$\Nc$ limit and may hint at the existence of a phase transition as a function of $\Nc$.
This would also deserve further investigation.

\begin{acknowledgments}
I am grateful to Kenji Fukushima, Larry McLerran, and Sanjay Reddy for useful discussions.
I thank Toru Kojo, Larry McLerran again, and Naoki Yamamoto for their comments on the manuscript.
I thank Ryan Abbott for the discussions and for providing me with the data of Ref.~\cite{Abbott:2023coj}.
Y.F.\ is supported by the Japan Society for the Promotion of Science (JSPS) through the Overseas Research Fellowship and by the INT's U.S. DOE Grant No. DE-FG02-00ER41132.
\end{acknowledgments}

\bibliographystyle{apsrev4-2}
\bibliography{gap}

\end{document}